\documentclass[structabstract]{aa}  
\usepackage{graphicx}
\usepackage{amsmath}
\usepackage{amssymb}
\usepackage{txfonts}               % A&A fonts
\usepackage{longtable,lscape}
\usepackage{hyperref}
\usepackage[authoryear]{natbib}                % improved management of bibliographic citations
\newcommand{\be}{\begin{eqnarray}}
\newcommand{\ee}{\end{eqnarray}}
\newcommand{\beq}{\begin{equation}}
\newcommand{\eeq}{\end{equation}}
\begin{document}
%

%\setcitestyle{authoryear}
   \title{A two-zone approach to neutrino production in gamma-ray bursts}

%  \subtitle{Overviewing the $\kappa$-mechanism}

\author{
  M.~M.~Reynoso\inst{1,2}
}

\institute{Department of Physics, University of Athens, Panepistimiopolis, GR 15783 Zografou, Greece \and Instituto de Investigaciones F\'{\i}sicas de Mar del Plata (CONICET - UNMdP), Facultad de Ciencias Exactas y Naturales, Universidad Nacional de Mar del Plata, Dean Funes 3350, (7600) Mar del Plata, Argentina}

   \date{Received November 29, 2013; accepted February 5, 2014}

% \abstract{}{}{}{}{} 
% 5 {} token are mandatory
 
  \abstract
   %{On Blazars and PKS}
   {Gamma-ray bursts (GRB) are the most powerful events in the universe. They are capable of accelerating particles to very high energies, so are strong candidates as sources of detectable astrophysical neutrinos.}
   {We study the effects of particle acceleration and escape by implementing a two-zone model in order to assess  the production of high-energy neutrinos in GRBs associated with their prompt emission.}
  {Both primary relativistic electrons and protons are injected in a zone where an acceleration mechanism operates and dominates over the losses. The escaping particles are re-injected in a cooling zone that propagates downstream.
The synchrotron photons emitted by the accelerated electrons are taken as targets for $p\gamma$ interactions, which generate pions along with the $pp$ collisions with cold protons in the flow. The distribution of these secondary pions and the decaying muons are also computed in both zones, from which the neutrino output is obtained.}
  % results heading (mapero nondatory)
  {We find that for escape rates lower than the acceleration rate, the synchrotron emission from electrons in the acceleration zone can account for the GRB emission, and the production of neutrinos via $p\gamma$ interactions in  this zone becomes dominant for $E_\nu>10^5$ GeV. For illustration, we compute the corresponding diffuse neutrino flux under different assumptions and show that it can reach the level of the signal recently detected by IceCube. }
  {}


   \maketitle
%
%________________________________________________________________

\section{Introduction}

Gamma-ray bursts (GRB) are intense and brief flashes of gamma rays that last from a fraction of a second to tens of seconds, releasing energies as high as $10^{51-53}${\rm erg} \cite{piran2004,meszaros2006}. {While short bursts with durations $t_{\rm GRB}^{\rm obs}\sim 2$ s are believed to be caused by the merger of compact stars in a binary system, long bursts ($t_{\rm GRB}^{\rm obs}\gtrsim 10$ s) are thought to be triggered by the collapse of a massive star into a black hole.} In the most accepted scenario, the prompt emission corresponding to the observed burst is supposed to come from synchrotron and/or inverse Compton emission of electrons that are accelerated in internal shocks of ejecta with various Lorentz factors, $\Gamma\sim 100-1000$ \cite[e.g.][]{rees1994,fenimore1996,kobayashi1997}. However, this is not the only possibility: photospheric models and acceleration by reconnection have also been proposed to explain such emission \cite[e.g.][]{meszaros2000,giannios2006,gao2012}.%(e.g. M\'{e}sz\'{a}ros \& Rees 2000; Gao et al. 2012).

Neutrino production in GRBs is expected if, for instance, protons are co-accelerated with the electrons responsible for the prompt emission. Then, $p\gamma$ and $pp$ interactions in the the baryon rich flow would lead to pion production, and thus to neutrinos \citep[e.g.][]{waxmanbahcall1997,guetta2004,murase2006}. In more recent studies, new calculations have been developed to obtain the possible neutrino flux under different assumptions %(e.g. H\"{u}mer et al. 2012; Murase et al. 2012; Baerwald et al. 2012; Vieyro et al. 2013).
\cite[e.g.][]{hummer2012,murase2012,baerwald2012,he2012}. It has also been proposed that in an earlier stage, while the jet is still propagating inside the collapsing star or just outside its surface, shocks may develop but without an observable photon counterpart, and only neutrinos would escape \cite{razzaque2004,ando2005,vieyro2013}. Another well studied possibility is the generation of neutrinos during the afterglow phase \cite[e.g.][]{waxman2000,dai2001}, which corresponds to a delayed low energy emission that occurs from hours to days after the prompt emission, and is commonly explained by external shocks with the interstellar medium. 

In the present work, we focus on the production of prompt neutrinos, considering the effects of a generic acceleration process acting on all charged particles, including the secondary pions and muons. To do this, we adopt a simple model with two zones: an acceleration zone and a cooling one, and we assume that the particles escaping from the former are injected into the latter. We find that in the cases where the escape rate is slower than the acceleration rate, then the synchrotron emission from the electrons in the acceleration zone can yield a flux that is consistent with GRB observations. Then, the co-accelerated protons can produce significant amounts of pions by $pp$ and $p\gamma$ interactions depending on the power injected in protons. The decaying muons can undergo acceleration in the cases of higher magnetic fields, for which their acceleration rate becomes higher than their decay rate. For illustration, we compute the diffuse neutrino flux that would be expected from GRBs under some different assumptions on the Lorentz factor and on the escape rate, and we compare these results with the the Waxman-Bahcall GRB flux \cite{waxmanbahcall1997} and with the data of the recent neutrino detection by IceCube \cite{aartsen2013}. 

This work is organized as follows. In Section 2, we describe the basic assumptions of the model, and in Section 3 we show the results obtained for the particle distributions: protons, electrons, pions and muons. In Section 4 we show illustrative results for the predicted broadband GRB photon flux, and in Section 5 we compute the corresponding diffuse fluxes of prompt neutrinos. The final comments are made in Section 6. 

%__________________________________________________________________

\section{Basics of the model}
%                                     Two column figure (place early!)
%______________________________________________ Gamma_1 (lg rho, lg e)

%--------------------- FIG 1
  \begin{figure}[htp]
   \centering
  \includegraphics[trim = 0mm 0mm 0mm 0mm, clip,width=0.6\linewidth,angle=0]{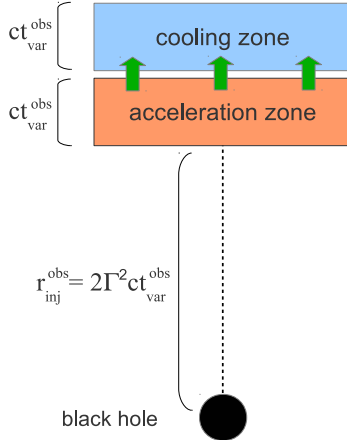}
   \caption{Basic elements of the model. See the text for details.}
              \label{fig1:sketch}%
    \end{figure}

The main components of the model are depicted in Fig. \ref{fig1:sketch}. The acceleration zone is the place where primary protons and electrons are injected and accelerated. The underlying idea is that a GRB of a total duration $t^{\rm obs}_{\rm GRB}\sim 10$ s is considered to be produced by many injection events ($\mathcal{N}_{\rm inj}=t^{\rm obs}_{\rm GRB}/t^{\rm obs}_{\rm var}$), each responsible for a peak of duration $ t^{\rm obs}_{\rm var}= 0.1- 0.01$s in the observed lightcurve. {Here, the superscripts $^{\rm obs}$ correspond to {a frame at the source location}, i.e., not corrected by redshift.}
The accelerated particles that escape are re-injected in a second zone where they lose energy.

The values of the basic parameters are estimated as in previous studies \cite[e.g.][]{piran2004,meszaros2006}. The distance from the central source to the initial position of the acceleration zone is related to the Lorentz factor of the flow $\Gamma$ and the variability time-scale as 
$$
r^{\rm obs}_{\rm inj}= 2 \Gamma^2 t^{\rm obs}_{\rm var} c
,$$
which{, in the case of $t^{\rm obs}_{\rm var}=0.01$ s,} yields $6\times 10^{12}$ cm and $5.4\times 10^{13}$ cm for $\Gamma=100$ and $\Gamma=300$, respectively.
 
The thickness of this zone in the comoving frame is 
$
\Delta r=  r^{\rm obs}_{\rm inj}/(2\Gamma)
$, and its comoving volume is $\Delta V= 4\pi r_{\rm obs}^2\Delta r$, where $r_{\rm obs}\simeq r^{\rm obs}_{\rm inj}+ c \, t/(2\Gamma)$ is the position of the acceleration zone as a function of the comoving time $t$. For simplicity, we assume that both the acceleration zone and cooling zone have the same volume and Lorentz factor.

We suppose that the bulk kinetic energy of the flow is $E_{\rm kin}=10^{52-53}$erg, so that the comoving number density of cold protons is given by
\beq
n_{\rm cold}= \frac{E_{\rm kin}}{\mathcal{N}_{\rm inj}\Gamma \Delta V m_pc^2}. 
\eeq
The magnetic energy is assumed to be a fraction $\epsilon_{B}$ of the kinetic energy, which implies a magnetic field \citep[e.g.][]{waxmanbahcall1997,murase2006,baerwald2012}:
%(e.g. Waxman \& Bahcall 1997, Murase \& Nagataki 2006, Baerwald et al. 2012):
\beq
B=\sqrt{\epsilon_{B} 8\pi m_p c^2 n_{\rm cold}}\nonumber.
\eeq
For instance, this yields $B=4.3\times 10^{5}$G for $\Gamma=100$ and $B=1.6\times 10^{4}$ G for $\Gamma=300$ if $\epsilon_{B}=0.1$.
In the acceleration zone, we suppose that there is a certain mechanism that increases the energy $E_i$ of particles of a type $i=\left\lbrace e,p,\pi,\mu \right\rbrace$ at a rate \cite[e.g.][]{begelman1990}
\be
t^{-1}_{\rm acc}(E_i)=  \frac{\eta \ e \ B \ c}{E_i}, \label{tacc}
\ee
where $\eta$ is an efficiency parameter. The relation between this acceleration rate and the rate of escape from acceleration zone the into the cooling zone will affect the energy dependence of the particle distributions \cite[e.g.][]{protheroestanev1999,drury1999,moraitis2007}. Here we assume that the escape rate is some fraction of the acceleration rate,
\be 
t_{\rm esc}^{-1}(E_i)= \xi_{\rm esc} t_{\rm acc}^{-1}(E_i).
\ee
A reference case is the one where $\xi_{\rm esc}=1$, which yields distributions proportional to $E^{-2}$ in the cooling zone \cite{kirk1998}. Still, in the present work we explore situations in which $\xi_{\rm esc}< 1$, since we are not specifying the nature of the acceleration mechanism and are exploring different cases in the context of GRBs. \footnote{We note, for example, that even cases with  $\xi_{\rm esc}\ll 1$ have been proposed in photospheric dissipative models \cite[e.g.][]{boschramon2012,drury2012,gao2012}.}

We next comment on the cooling processes, and we leave the question of the particle distributions and the method of calculation for Section 3.

%------------------------------------------gamma ray  bursts are originated----------------- FIG 2 tep
   \begin{figure*}
   \centering
\includegraphics[trim = 0mm 0mm 0mm 0mm, clip,width=0.9\linewidth]{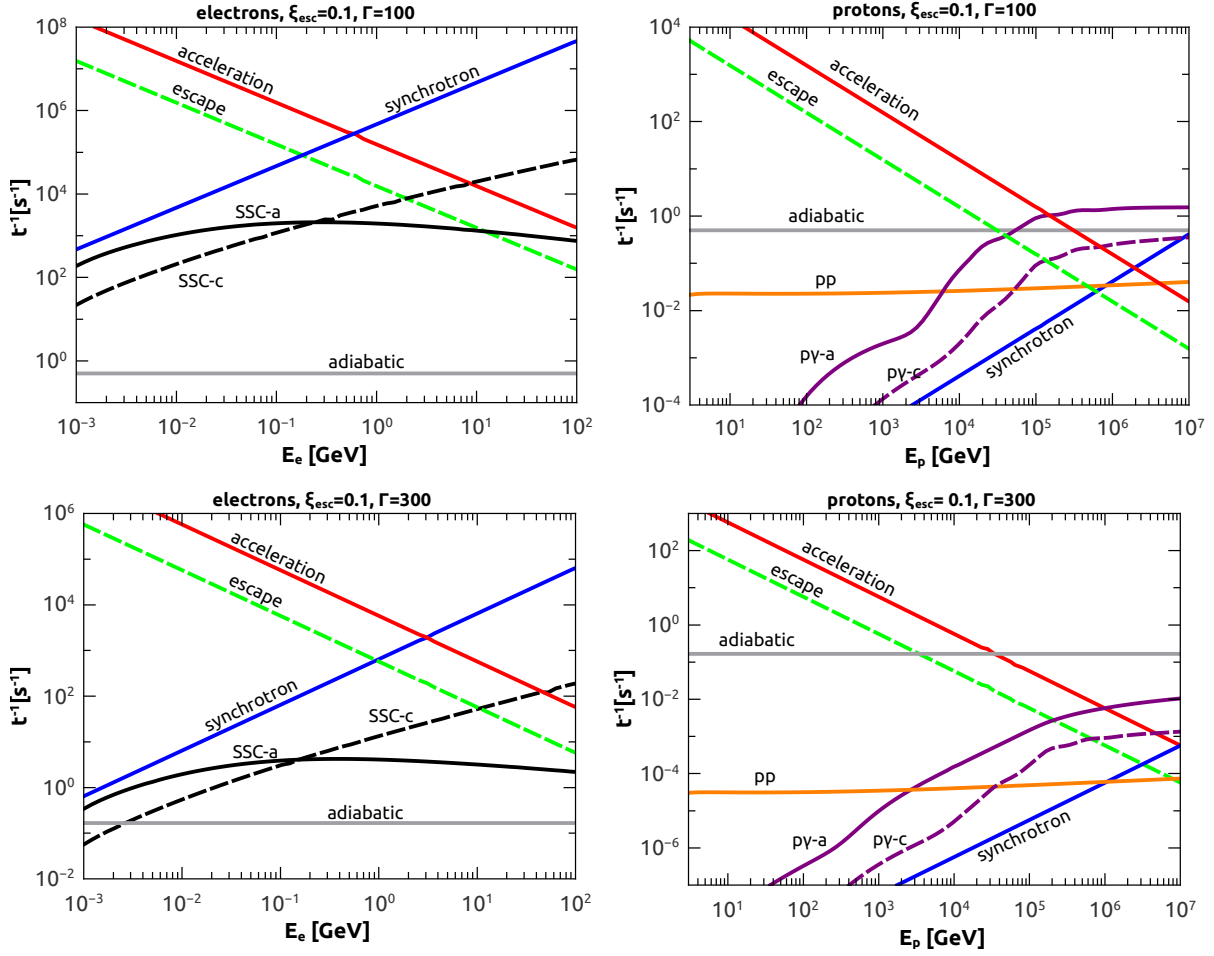}
      \caption{Acceleration and cooling rates for electrons and protons for a bulk Lorentz factor $\Gamma=100$ and $300$ in the top and bottom panels, respectively. The labels SSC-a(c) and $p\gamma$-a(c) indicate the respective processes for the acceleration (cooling) zone. The escape rate shown is $10^{-1}t^{-1}_{\rm acc}$.  
              }
         \label{fig2:tep}
   \end{figure*}
%
%___________________________________________________________________

%----------------------------------------------------------- FIG 3 tpimu
   \begin{figure*}
   \centering
\includegraphics[trim = 0mm 0mm 0mm 0mm, clip,width=0.9\linewidth]{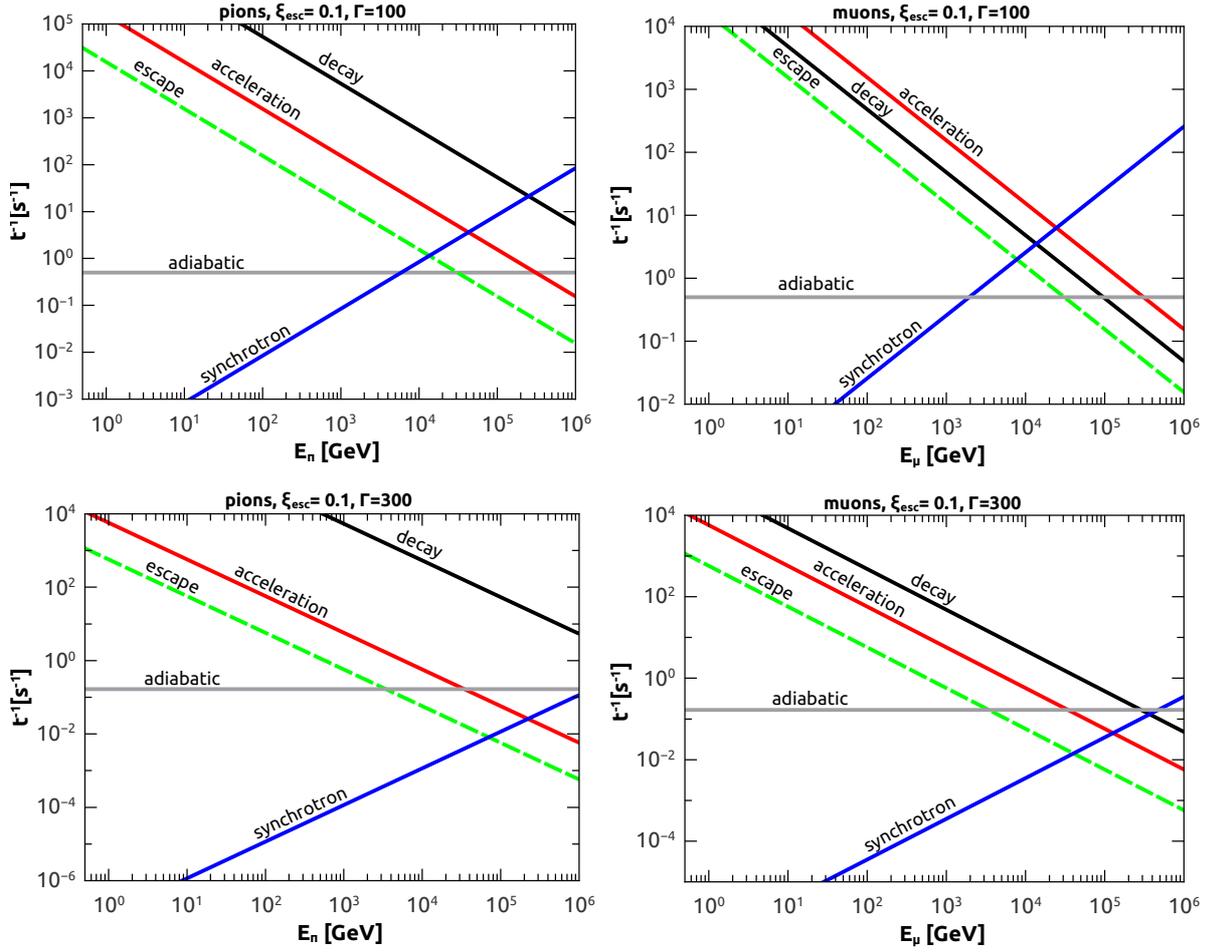} 
      \caption{Rates of acceleration, escape, decay, and cooling for pions and muons. The top panels correspond to     
  $\Gamma=100$ and the bottom ones to $\Gamma=300$.      }
         \label{fig3:tpimu}
   \end{figure*}
%
%___________________________________________________________________

\subsection{Cooling processes}

The synchrotron energy loss rate{, in the CGS system of units, is}
 \beq
{t}_{\rm sync}^{-1}(E_i)=\frac{4}{3}\left(\frac{m_e}{m_i}\right)^3\frac{
\sigma_{\rm T}{B}^2}{m_e c \ 8\pi}\frac{E_i}{m c^2}\label{tsyn}.
 \eeq
We consider an adiabatic cooling with a rate similar to the inverse of the dynamical timescale \cite[e.g.][]{murase2006},
\beq
 {t}_{\rm ad}^{-1}(t)\approx \Gamma \frac{c}{r_{\rm obs}}= \frac{1}{t}  
\eeq
where $t$ is the comoving time.

To compute the inverse Compton cooling rate, a soft photon field is necessary as a target for the electrons. We assume that these soft photons are mainly due to the synchrotron radiation of the same electron population, which have a differential density (in units of ${\rm energy}^{-1}{\rm length}^{-3}$)
 \beq
n_{\gamma}^{(e-\rm{syn})}(E_{\gamma},t)= \left(\frac{\Delta r}{c}\right) 4\pi Q_\gamma^{(e-\rm{syn})}(E_\gamma,t),
 \eeq
where the synchrotron emissivity, in units of (${\rm energy}^{-1}{\rm length}^{-3}{\rm sr}^{-1}{\rm time}^{-1}$), is
\begin{multline}
%Q_{\gamma,{\rm syn}}(E_\gamma)=   \frac{1}{E_\gamma}\int_{m_e c^2}^{\infty} dE' 4\pi  P_{\rm syn}
%    N_{e}(E').\label{varepsilon}
Q_{\gamma}^{(e-{\rm syn})}(E_\gamma,t)=   \frac{\sqrt{2} e^3 B}{m_e c^2 h}
\frac{4\pi}{E_{\rm cr}}\int_{m_e c^2}^{\infty} dE' \\
  \times \int_{E_{\gamma}/E_{\rm cr}}^\infty d\zeta
K_{5/3}(\zeta)
    N_{e}(E',t).\label{varepsilon}
\end{multline}

Here, $N_e$ is the electron distribution in units of (${\rm energy}^{-1}{\rm length}^{-3}$), $K_{5/3}(\zeta)$ is the modified Bessel function of order $5/3$, and {the critical energy, close to the peak of the synchrotron spectrum, is \citep[e.g.][]{blumenthal1970}}
$$E_{\rm cr}= \frac{\sqrt{6}he B}{4\pi m_e c}\left(\frac{E'}{m_e
c^2}\right)^2 $$
{in the comoving frame.}

The synchrotron self-Compton cooling rate is then approximated by \citep[e.g.][]{jones1968} %(e.g. Jones 1968)
 \begin{multline}
 {t}^{-1}_{\rm SSC}(E_e,t)= \frac{3 m_e^2 c^5 \sigma_{\rm T}}{4 {E}^3}\int_{{E}_{\rm ph}^{\rm(min)}}^{E_e}dE_{\rm ph } \frac{n_{\gamma}(E_{\rm ph},t)}{{E}_{\rm ph}} \\ \times \int_{{E}_{\rm ph}}^{\frac{\Gamma_e}{\Gamma_e+1}E_e} dE_\gamma  F(q)\left[E_\gamma-E_{  \rm ph}\right], \label{tIC}
 \end{multline}
where {${E}_{\rm ph}^{\rm(min)}$ is the lowest energy of the available background of synchrotron photons, and}
 \be
 F(q)= 2q \ln q + (1 + 2q) (1 - q)+ \frac{1}{2}(1 - q)\frac{(q \Gamma'_{\rm e})^2}{1 + \Gamma'_{\rm e} },
 \ee
with $\Gamma_{\rm e}=4 E_{\rm ph}E_e/(m_e^2c^4)$ and
 $q={E_\gamma}\left[{\Gamma_{\rm e}E_{\rm ph}(1-{E_\gamma/}{E_{\rm ph}})}\right]^{-1}.$

As for protons, the $p\gamma$ cooling rate is 
\begin{multline}
{t}_{p\gamma}^{-1}(E_p,t)=\frac{c}{2\gamma_p^2}\int_{\frac{\epsilon_{\rm
th}^{(\pi)}}{2\gamma_p}}^{\infty} dE_{\gamma}\frac{n_{\gamma}(E_{\gamma},t)}{{E}_{\gamma}^2}  \\
\times \int_{\epsilon_{\rm th}^{(\pi)}}^{2E_{\gamma}\gamma_p} d \epsilon_{\rm r}
  \sigma_{p\gamma}^{(\pi)}(\epsilon_{\rm r})K_{p\gamma}^{(\pi)}(\epsilon_{\rm r})
  \; \epsilon_{\rm r}, \label{tpg}
\end{multline}
where, $\epsilon_{\rm th}^{(\pi)}=150$ MeV, and we use the expressions for the cross section $\sigma_{p\gamma}^{(\pi)}$
and the inelasticity $K_{p\gamma}^{(\pi)}$ given in Atoyan \& Dermer (2003). The $e^+e^-$ pair production by $p\gamma$ collisions (Bethe-Heitler process) was also included as in the $t_{p\gamma}^{-1}$ following Begelman et al. (1990).

The energy loss rate due to inelastic $pp$ collisions is
 \be
{t}_{pp}^{-1}(E_p)= {n_{\rm cold}} \; c \; \sigma_{pp}^{\rm(inel)}(E_p)K_{pp},
 \ee
where the inelasticity coefficient is $K_{pp}\approx 1/2$ and the corresponding cross section can be approximated as in Kelner et al. (2006).

For the parameter values of Table \ref{table1}, we show in Fig. \ref{fig2:tep} the acceleration and cooling rates for electrons and protons, separating the cases of $\Gamma= 100$ and $\Gamma=300$ in the upper and lower panels, respectively. {Here we note that the rates are much lower for $\Gamma=300$ than for $\Gamma=100$, which will require more energy to be injected in the former case in order to have similar radiative outputs, as we will see below}. The escape rate is also shown, which in this case is a fraction $\xi_{\rm esc}=0.1$ of the acceleration rate as mentioned above. The acceleration and cooling rates for pions and muons are shown in Fig. \ref{fig3:tpimu}, where the corresponding decay rates are also included.

 \begin{table*}
     \caption[]{Parameters of the model}\label{table1}
{\small
$$  \begin{tabular}{|c|c|}
    \hline % after \\: \hline or \cline{col1-col2} \cline{col3-col4} ...    
    Parameters &  values$^\dagger$     \\
    \hline
    $E_{\rm GRB} $: total energy of GRB photons & $10^{53}{\rm erg}$\\    
    $t^{\rm obs}_{\rm GRB} $: total GRB duration & $10$ s\\
    $t^{\rm obs}_{\rm var} $: variability timescale & $0.01$ s\\    
    $\gamma_{\rm inj} $: Lorentz factor of injection & $10$ \\        
    $\eta $: particle acceleration efficiency & $4\times 10^{-5}$ \\                
    $\xi_{\rm esc} $: $t^{-1}_{\rm esc}/t^{-1}_{\rm acc}$, ratio of escape to acceleration rates & $0.25$ and $0.1$ \\            
    $\epsilon_{B} $: fraction of energy in $B$ & $0.1$\\
    $\Gamma $: bulk Lorentz factor & $100$; $300$\\
$L_e$: comoving power injected in electrons & $\sim 4\times 10^{44}{\rm erg/s}; \sim 10^{43}{\rm erg/s}$\\
$L_p$: comoving power injected in protons & $\sim 3\times 10^{43}{\rm erg/s}; \sim 5\times 10^{45}{\rm erg/s}$ \\
 $\Delta E_e$: total energy in electrons (Eqs. 20-22) & $\sim 2\times 10^{47}$erg; $\sim  10^{47}$ erg \\
 $\Delta E_p$: total energy in protons (Eqs. 20-22)& $\sim 3\times 10^{46}$erg; $\sim  10^{49}$ erg\\
       \hline	
      \end{tabular}$$
}$$\mbox{$^\dagger$Values corresponding to $\Gamma=100$ and $\Gamma=300$ appear separated by ``;".}$$
 \end{table*}

\section{Particle distributions}

To obtain the distribution of each particle type $i=\left\lbrace e,p,\pi,\mu \right\rbrace$ in each zone, we solve the general kinetic equation
 \beq
\frac{\partial N_{i}}{\partial t} +\frac{\partial\left[\dot{E}_{i} N_{i,}\right]}{\partial E_i}
+ \frac{N_{i}}{t_{\rm esc}}=  Q_i(E_i,t) \label{kineticeq},
 \eeq
where $Q_i$ is the injection term in units of $({\rm energy}^{-1}{\rm length}^{-3}{\rm time}^{-1})$, and the energy change is 
$\dot{E}_i(E_i,t)\equiv \frac{dE_i}{dt}$. The escape term is only present for particles in the acceleration zone, and in the case of pions and muons, the term of decay $(N_i t_{\rm dec}^{-1})$ must be also added in the left-hand side.

For the primary electrons and protons ($i=\{e,p\}$) in the acceleration zone, where acceleration is assumed to operate during the time of the injection event $t_{\rm var}= 2\Gamma t^{\rm obs}_{\rm var}$ in the comoving frame, we use a mono-energetic injection
\be
   Q_i(E_i,t)= K_i \ H(t-t_0)\ H(t_0+ t_{\rm var}-t) \ \delta(E_i- m_ic^2\gamma_{\rm inj}).
\ee
Here $H$ is the Heaviside step function, $\gamma_{\rm inj}$ the Lorentz factor of the injected particles, and $K_i$ a normalization constant
\beq
K_i= \frac{L_i}{\Delta V \  m_ic^2\gamma_{\rm inj}}.  
\eeq
This one is fixed by the corresponding power injected in the comoving frame during the time $t_{\rm var}$:
\beq
L_i= q_i L_{\rm GRB}=q_i \frac{E_{\rm GRB}}{\mathcal{N}_{\rm inj} \Gamma \ t_{\rm var}}. \label{qeqp}
\eeq
{For instance, with the parameter values of Table \ref{table1} we obtain $L_{\rm GRB}= \lbrace 5\times 10^{47}{\rm erg}~{\rm s}^{-1}; \ 5.5\times 10^{46}{\rm erg}~{\rm s}^{-1}\rbrace$ for $\Gamma=100$ and $\Gamma=300$, respectively. The corresponding values of $q_i$ are those that yield the values of $L_i$ appearing in Table \ref{table1}: $q_e\sim 8\times 10^{-4}$ and $q_p\sim 6\times 10^{-5}$ for $\Gamma=100$, and  $q_e\sim 2\times 10^{-4}$ and $q_p\sim 0.1$ for $\Gamma=300$.}

%where $L_{\rm GRB}= {E_{\rm GRB}}/(\mathcal{N}_{\rm inj} \Gamma \ t_{\rm var}).$

The energy change in the acceleration zone includes both acceleration and losses, 
\beq
   \dot{E}_i(E_i)=  E_i\times \left[t^{-1}_{\rm acc}(E_i) -t^{-1}_{i,\rm{loss}}(E_i) \right],
\eeq
and the solution can be found using the method of the characteristics, through

\beq 
N_i(E_i,t)= \int_{t_0}^t dt'Q_i(E',t')\exp\left[-\int_{t'}^{t} 
 dt''\left(\frac{\partial \dot{E}_i}{\partial E_i}- t^{-1}_{\rm esc} \right)  \right], \label{Solformal}
\eeq 
which has units of (${\rm energy}^{-1}{\rm length}^{-3}$). {In the case of pions and muons, the effect of decay is included by replacing $t^{-1}_{\rm esc}\rightarrow t^{-1}_{\rm esc}+t^{-1}_{\rm dec}$.}

The process of calculation of the different particle distributions is as follows. We start by computing the electron distribution in the acceleration zone, $N^{\rm acc}_e(E_e,t)$, taking acceleration and synchrotron cooling into account. As can be seen in Fig. \ref{fig2:tep}, adiabatic  cooling for electrons is not important, and neither is the synchrotron self-Compton cooling at high energies for typical GRB parameters (see Table \ref{table1}). We then compute the proton distribution $N^{\rm acc}_p(E_p,t)$ considering acceleration, adiabatic cooling, synchrotron cooling, $pp$ interactions, and $p\gamma$ interactions with the synchrotron photons of electrons as targets. In the third place, we compute the injection of pions $Q^{\rm acc}_\pi(E_\pi,t)$ by both $pp$ and $p\gamma$ collisions and use it in the right-hand side of Eq. (\ref{kineticeq}) to obtain the distributions of pions in the acceleration zone, $N^{\rm acc}_\pi(E_\pi,t)$. After this, we compute the injection of muons $Q^{\rm acc}_\mu(E_\mu,t)$ and obtain $N^{\rm acc}_\mu(E_\mu,t)$. 

Particles escaping from the acceleration zone are re-injected in the cooling zone, with an injection $(N_i^{\rm acc}(E_i,t)\ t^{-1}_{\rm esc})$. We solve Eq. (\ref{kineticeq}) without any acceleration to obtain each $N_i^{\rm cool}(E_i,t)$ following the same order as for the acceleration zone.

\begin{figure*}
   \centering
   %                    l   b   r   t
\includegraphics[trim = 0mm 0cm 0mm 0mm, clip,width= \linewidth]{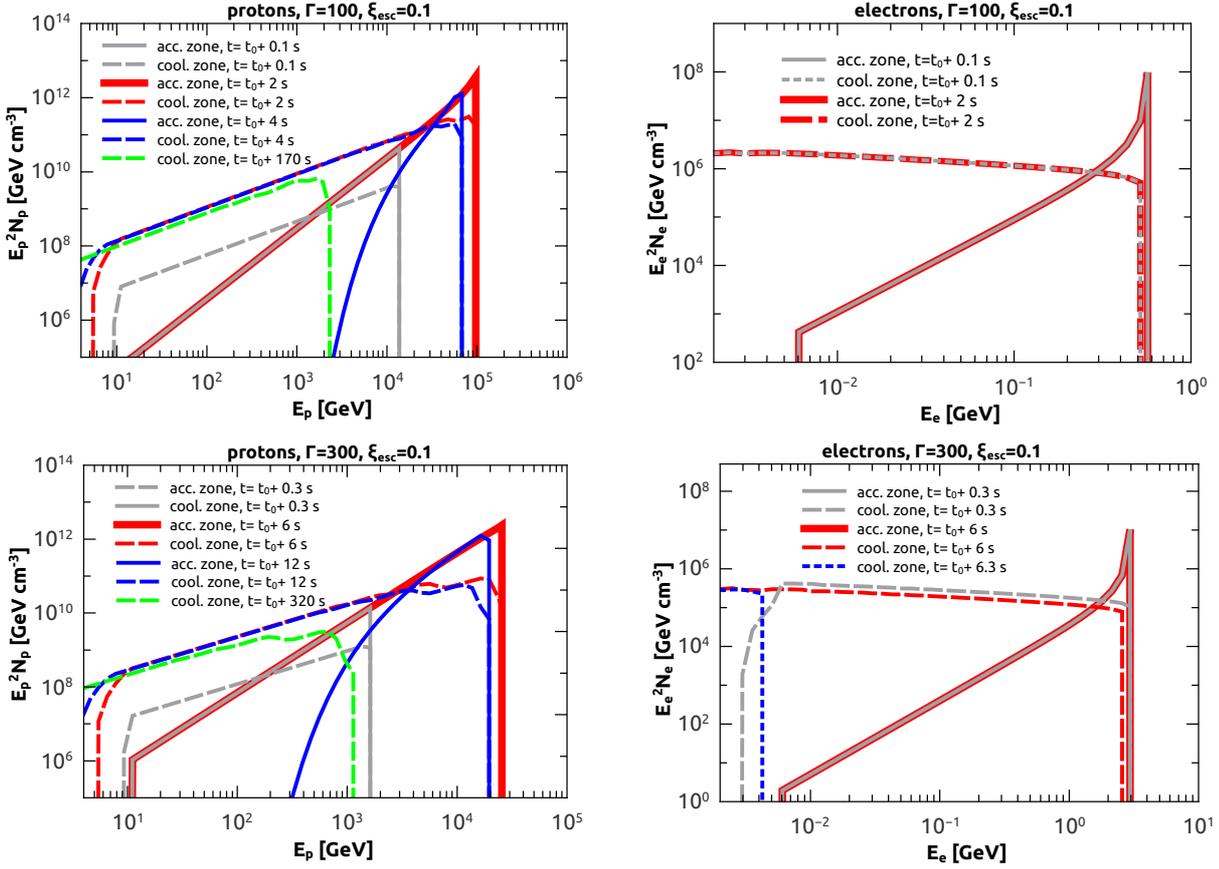}  %{Nep_g100_g300_v2.eps}
      \caption{Distributions of electrons and protons multiplied by the squared energy and evaluated at different times corresponding to the acceleration zone (solid lines) and to the cooling zone (dashed lines), for $\Gamma=100$ (top panels) and $\Gamma=300$ (bottom panels). 
              }
         \label{fig4:Nep_g100_g300}
   \end{figure*}

The most important cooling mechanisms for all particle types are synchrotron and adiabatic cooling, so that we can write a characteristic equation:
\be
\frac{dE_i}{dt}= -\frac{E_i}{t}- b_i E_i^2,
\ee
where the first term is the adiabatic energy loss and the second term is the synchrotron energy loss assuming for simplicity a constant magnetic field, with $$b_i=\frac{4}{3}\left(\frac{m_e}{m_i}\right)^3 \frac{c \sigma_T B^2}{8\pi m_e c^2}\frac{1}{m_i c^2}.$$
The solution can be found as in Kardashev (1962), using the characteristic curve that gives the energy $E'>E$ for early times $t'<t$,
\be
E'_i(t';E,t)= \frac{E}{b_i E_i \ t \ \log\left(\frac{t'}{t}\right)+ \frac{t'}{t}},
\ee
and substituting in Eq. (\ref{Solformal}).
In Fig. \ref{fig4:Nep_g100_g300}, we show the distributions of electrons and protons evaluated at different times in the acceleration and in the cooling zone. {The initial times used are $t_0=2\Gamma t^{\rm obs}_{\rm var}=t_{\rm var}=\lbrace 2 \ {\rm s},6 \ {\rm s}\rbrace$ for $\Gamma=\lbrace 100,300\rbrace$, which also correspond to the injection periods in each case.} Pile-ups occur at the maximum energy where the acceleration rate equals the cooling one. Although a divergence appears at exactly this energy, the injected particles never reach it because there is a time limitation given by the duration of injection. In the case of electrons, the synchrotron cooling is so fast that once the injection is switched off, there are no more high energy electrons in both zones, while the protons take a much longer time to lose their energy. {We note that the distributions in the cooling zone are steeper than those in the acceleration zone because of the $\propto E_i^{-1}$ dependence assumed for the escape rate, and this gives a high density distribution for low energies in the cooling zone, as compared to the acceleration zone. }

As for the energy involved, in addition to the power injected in mono-energetic electrons and protons ($L_i$, see Table \ref{table1}), particles undergo acceleration up to a maximum energy, experience losses, and a fraction of them can escape to the cooling zone. As a result, the total energy that is given to the particles can be computed as $\Delta E_i= \Delta E_i^{\rm acc}+\Delta E_i^{\rm esc}+ \Delta E_i^{\rm loss}$, with
\be
\Delta E_i^{\rm acc} &=& \Delta V \left(\int_{\gamma_{\rm inj}m_ic^2}^{E_i^{\rm max}} dE_i~E_i~N_i^{\rm acc}(E_i,t_0+t_{\rm var} )   \right) \label{DeltaEacc}\\
\Delta E_i^{\rm esc} &=& \Delta V \left(\int_{t_0}dt\int_{\gamma_{\rm inj}m_ic^2}^{E_i^{\rm max}} dE_i~E_i~ t^{-1}_{\rm esc}(E_i)~{N_i^{\rm acc}(E_i,t)}   \right)\label{DeltaEesc}\\
\Delta E_i^{\rm loss} &=& \Delta V \left(\int_{t_0}dt\int_{\gamma_{\rm inj}m_ic^2}^{E_i^{\rm max}} dE_i~E_i~ t^{-1}_{i,\rm{loss}}(E_i)~{N_i^{\rm acc}(E_i,t)}   \right)\label{DeltaEloss}.
\ee
In the cases studied, we obtain 
{ the values shown in Table \ref{table1} for $\Gamma=100$ and $\Gamma=300$, which}
are found to produce a similar level of radiation and neutrinos as we see below. The main difference is that protons in the case of $\Gamma=300$, a much higher energy has to be given to the proton population because their cooling efficiency is much lower than in the case with $\Gamma=100$.

%----------------------------------------------------------- FIG 4 Nep_g100g300
%___________________________________________________________________
% \subsection{Production of secondary  particles}

We now describe how we compute the injection of the secondary particles.
\subsection{Pions}
Proton interactions with protons and low energy photons produce pions. The pion injection due to $pp$ collisions is is calculated as
\begin{multline}
Q_{\pi, pp}(E_\pi,t)=n_{\rm cold} c\int_{0}^{1} \frac{dx}{x} N_p \left(\frac{E_\pi}{x},t\right)
%\\ \times
F_\pi\left(x,\frac{E_\pi}{x}\right)\sigma_{pp}^{\rm(inel)} \left(\frac{E_\pi}{x}\right)
\label{Qpipp}
\end{multline}
where the distribution of pions produced per $pp$ collision is \cite{kelner2006}
\begin{multline}
F_\pi\left(x,\frac{E_\pi}{x}\right)=4\alpha B_\pi
x^{\alpha-1}\left(\frac{1-x^\alpha}{1+r'
x^\alpha(1-x^\alpha)}\right)^4 \\ \times \left(
\frac{1}{1-x^\alpha}+ \frac{r'(1-2x^\alpha)}{1+
r'x^\alpha(1-x^\alpha)} \right) \left(1-\frac{m_\pi c^2}{
E_\pi}\right)^{1/2},
\end{multline}
with $x=E_\pi/E'$, $ B_\pi=a'+ 0.25$, $a'= 3.67+ 0.83 L+ 0.075 L^2$, $r'=
2.6/\sqrt{a'}$, and $\alpha= 0.98/\sqrt{a'}$.

Similarly, the injection of charged pions produced by $p\gamma$ interactions is
\begin{multline}
Q_{\pi, p\gamma}(E_\pi,t)= \int_{E_\pi} dE_p N_p(E_p,t) \;
{\omega}_{p\gamma}^{(\pi)}(E_p) %\\ \times
\mathcal{N}_\pi(E_p,t) \; \delta\left(E_\pi- \frac{E_p}{5}\right) \\
= 5 \; N_p(5E_\pi,t) \; {\omega}^{(\pi)}_{p\gamma}(5E_\pi) \;
\mathcal{N}_\pi(5E_\pi).\label{Qpipg}
\end{multline}
Here, ${\omega}_{p\gamma}^{(\pi)}$ is the $p\gamma$ collision frequency defined as \citep[][]{atoyandermer2003} %(Atoyan \& Dermer 2003),
%\begin{multline}
\beq
 {\omega}_{p\gamma}^{(\pi)}(E_p,t)= \frac{c}{2\gamma_p^2} \int_{\frac{\epsilon_{\rm
th}^{(\pi)}}{2\gamma_p}}^{\infty} dE_{\gamma}\frac{n_{\gamma}(E_{\gamma},t)}{{E_{\gamma}}^2}
%\\  \times 
\int_{\epsilon_{\rm
th}^{(\pi)}}^{2\epsilon\gamma_p} d \epsilon
  \sigma_{p\gamma}^{(\pi)}(\epsilon) \epsilon,
\eeq  
% \end{multline}
and the mean number of $\pi^+$'s or $\pi^-$'s is approximately
 \be
\mathcal{N}_\pi \approx \frac{p_1}{2} + 2p_2.
 \ee
This number depends on the probabilities of single pion and multi-pion production $p_1$ and $p_2=1-p_1$. Given that  the mean inelasticity function is $\bar{K}_{p\gamma}={t_{p\gamma}^{-1}/\omega_{p\gamma}^{(\pi)}}$, we have  
\be
p_1=\frac{K_2-\bar{K}_{p\gamma}}{K_2-K_1}, \ee where $K_1=0.2$ and
$K_2=0.6$.
%
%----------------------------------------------------------- FIG 5 Npimu g100 g300
   \begin{figure*}
   \centering
\includegraphics[trim = 0mm 0mm 0mm 0mm, clip,width=0.9\linewidth]{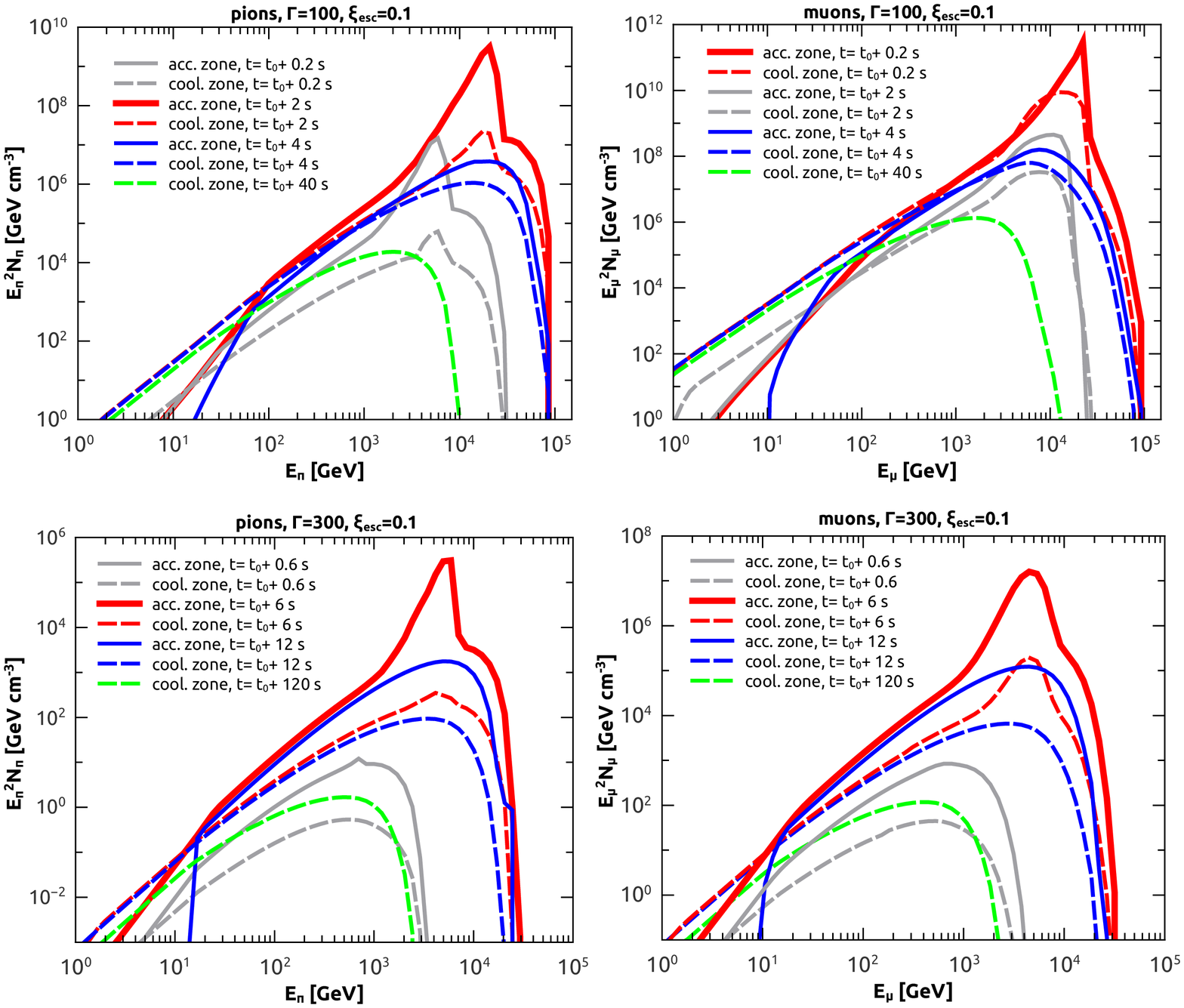} %{Npimu_g100_g300.eps} 
      \caption{Distributions of pions and muons multiplied by the squared energy and evaluated at different times corresponding to the acceleration zone (solid lines) and to the cooling zone (dashed lines), for $\Gamma=100$ (top panels) and $\Gamma=300$ (bottom panels).    
              }
         \label{fig5:Npimu_g100_g300}
   \end{figure*}
%___________________________________________________________________
\subsection{Muons}
The injection of muons is treated following Lipari et al. (2007), considering left-handed and right-handed muons separately with their decay spectra:
\be \frac{dn_{\pi^- \rightarrow
\mu^-_L}}{dE_\mu}(E_\mu;E_\pi)= \frac{r_\pi(1-x)}{E_\pi
x(1-r_\pi)^2}\ H(x-r_\pi) \\
\frac{dn_{\pi^- \rightarrow \mu^-_R}}{dE_\mu}(E_\mu;E_\pi)=
\frac{(x-r_\pi)}{E_\pi x(1-r_\pi)^2} \ H(x-r_\pi),
\ee
where $x= E_\mu/E_\pi$ and $r_\pi= (m_\mu/m_\pi)^2$.

Assuming CP invariance implies $dn_{\pi^- \rightarrow\mu^-_L}/dE_\mu= dn_{\pi^+ \rightarrow \mu^+_R}/dE_\mu$, and since the total distribution obtained for all charged pions is
$N_\pi= N_{\pi^+}+ N_{\pi^-}$, the injection of left handed muons is
%\begin{multline}
\beq
  Q_{\mu^-_L,\mu^+_R}(E_\mu,t)= \int_{E_\mu}^{\infty} dE_\pi 
%  \\ \times
  \frac{N_{\pi}(E_\pi,t)}{T_{\pi,{\rm dec}}(E_\pi)} \frac{dn_{\pi^- \rightarrow
  \mu^-_L}}{dE_\mu}(E_\mu;E_\pi). \label{QmuL}
 \eeq  
%\end{multline}
Similarly, the injection of right handed muons is
%\begin{multline}
\beq
  Q_{\mu^-_R,\mu^+_L}(E_\mu,t)= \int_{E_\mu}^{\infty} dE_\pi 
%  \\ \times
  \frac{N_{\pi}(E_\pi,t)}{T_{\pi,
{\rm dec}}(E_\pi)} \frac{dn_{\pi^- \rightarrow
  \mu^-_R}}{dE_\mu}(E_\mu;E_\pi). \label{QmuR}
\eeq  
%\end{multline}

In Fig. \ref{fig5:Npimu_g100_g300}, we show the obtained distributions of pions and muons in both zones at different times. The bumps in the distributions at energies around $\sim 10^4$ GeV is due to the contribution of $p\gamma$ interactions that becomes greater than that of $pp$. In the particular case of muons for $\Gamma=100$, the peak is more pronounced because muons are undergoing acceleration, and the maximum energy where acceleration equals losses is also around $\sim 10^4$ GeV. In the cooling zone, the pion injection is dominated by the produced by $p\gamma$ and $pp$ interactions in this zone, since the decay rate is much greater than the escape rate for the cases studied here, so pions generated in the acceleration zone will decay there, as can be seen in Fig. \ref{fig3:tpimu}. For muons, the decay is slower, so for $\xi_{\rm esc }>0.1$, the escape rate approaches the decay rate, and there is a non-negligible injection of escaping muons into the cooling zone, which is added up to the contribution coming from pions created in the cooling zone itself.

\section{Electromagnetic emission}\label{sec:photons}
%----------------------------------------------------------- FIG 6 E2dfluga g100
   \begin{figure*}
   \centering
\includegraphics[trim = 0mm 0mm 0mm 0mm, clip,width=0.9\linewidth]{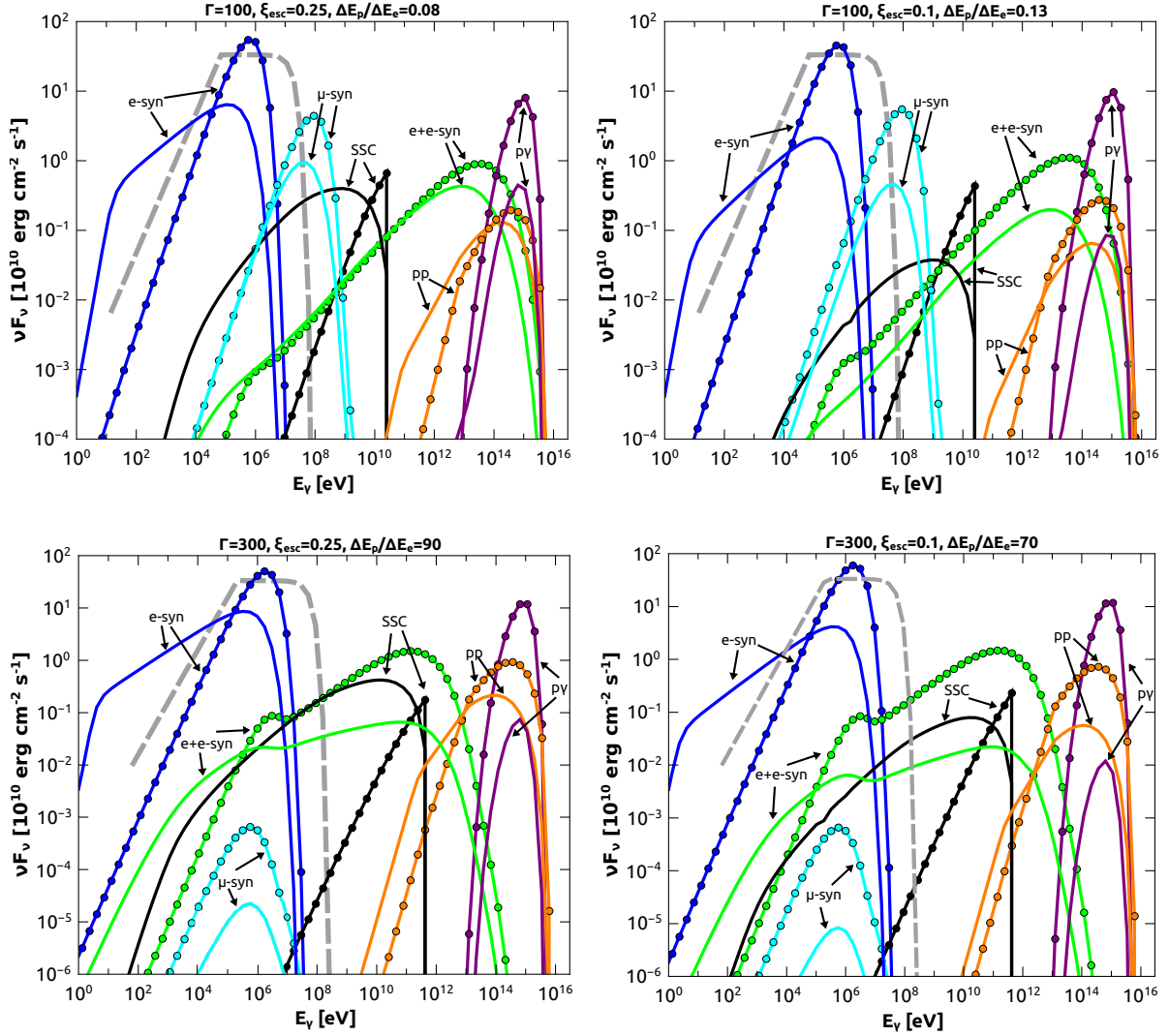} %{E2dfluga_g100_g300_noGG.eps} %{sedjetbw.eps}
      \caption{SED of photons obtained for a GRB with $\Gamma=100$ (top panels) and with $\Gamma=300$ (bottom panels) originated in both the acceleration zone (solid lines with circles) and in the cooling zone at different times (solid lines). A typical GRB photon field is shown for comparison in a grey thick dashed line. The dominant process are included: electron synchrotron (blue), inverse Compton (black), muon synchrotron (cyan), secondary $e+e-$ synchrotron (green), $p\gamma$ (purple), and $pp$ (orange). The different values adopted for $\xi_{\rm esc}$ are indicated.}
         \label{fig6:E2dfluga_g100_g300}
   \end{figure*}
%___________________________________________________________________
%----------------------------------------------------------- FIG 7 E2dflunu g100 g300
   \begin{figure*}
   \centering
\includegraphics[trim = 0mm 0mm 0mm 0mm, clip,width=0.9\linewidth]{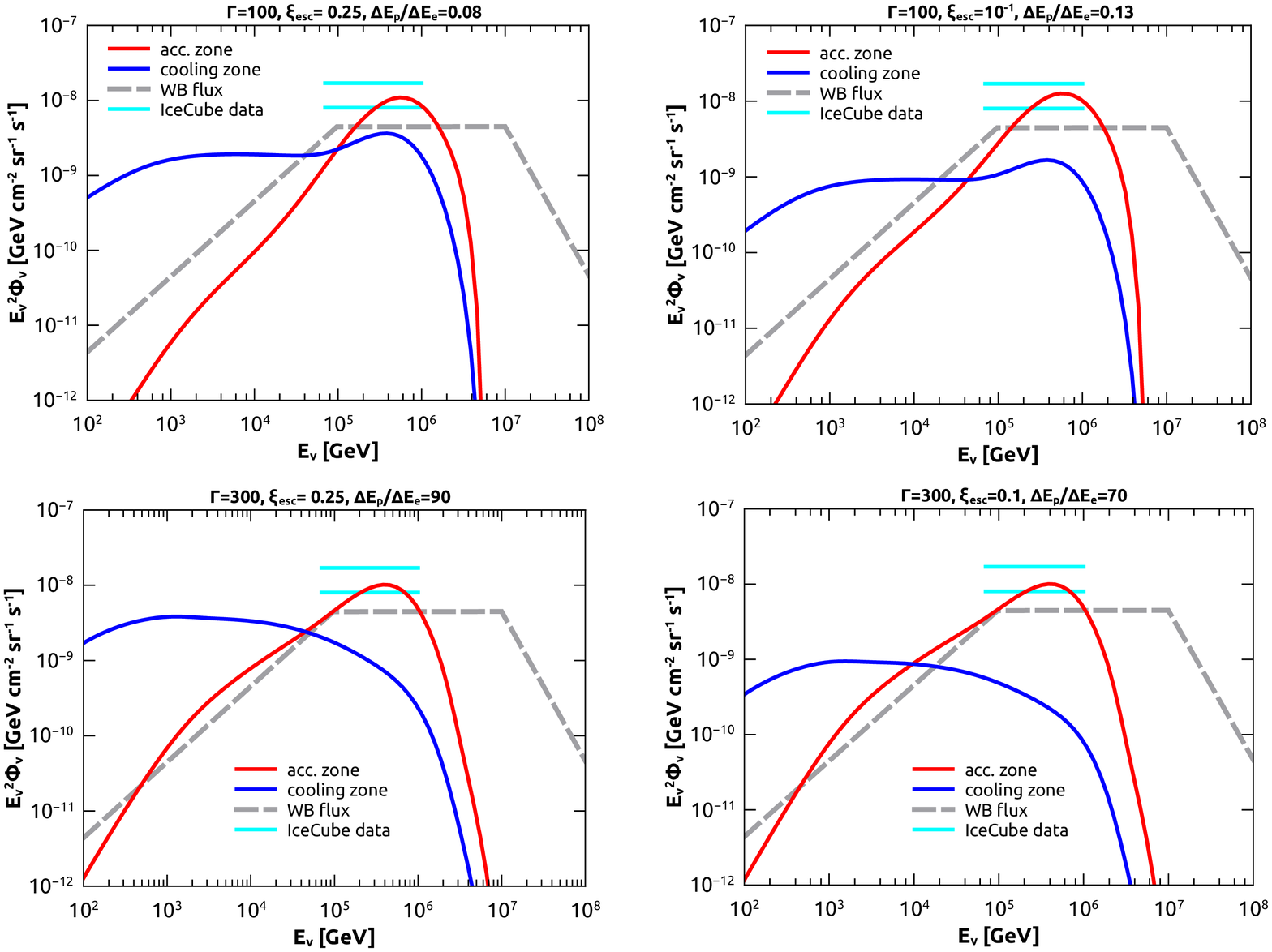} %{E2dflunudiff.eps}
      \caption{Diffuse flux of muon neutrinos predicted for GRBs, associated to their prompt emission in the case of $\Gamma=100$ (top panels) and $\Gamma= 300$ (bottom panels) for $\xi_{\rm esc}=0.25$ and $\xi_{\rm esc}=0.1$ in the left and right panels, respectively. The contribution from the acceleration zone is marked in red and the one from the cooling zone in blue. The reference Waxman-Bahcall flux is also shown for comparison.             }
         \label{fig7:Esdflunu_g100_g300}
   \end{figure*}
Here we present results for the broadband photon emission produced by the different particle populations in both zones of the present model. We chose two cases in which the escape rate is slower than the acceleration rate since this can give rise to significant synchrotron emission of electrons in the acceleration zone. The peak of this emission is related to the maximum energy of the electrons, which depends on the magnetic field and on the efficiency of the acceleration $\eta$. We find that this peak can fall within the correct energy range and intensity as turns out when we compare it with a usually adopted spectrum for GRBs, the following broken power law \cite[e.g.][]{murase2006,lipari2007,baerwald2012}:
\be 
n_{\rm bpl}(E_\gamma)= C_{\gamma}\left\lbrace \begin{array}{cc}
  \left(\frac{E_\gamma}{E_{\rm break}}\right)^{-1}   \hspace{1 cm} {\rm for \ } 0.2 \ {\rm eV}<E_\gamma<1 {\rm keV} \\
  \left(\frac{E_\gamma}{E_{\rm break}}\right)^{-2}
        \exp{\left(-\frac{E_\gamma}{300\ {\rm keV}}\right)} \hspace{1 cm} {\rm for \ } E_\gamma \geq \ {\rm keV}.

\end{array} \right.
\ee
Here, the constant $C_\gamma$ is fixed by specifying the energy density of these population of photons, which we take to be equal to the magnetic energy density, as assumed in the works mentioned.

In Fig. \ref{fig6:E2dfluga_g100_g300} we show the spectral energy density (SED) of photons corresponding to this broken power law profile, and we include all the relevant contributions within our model arising from the processes in both zones evaluated at the time of maximum emission, $t=t_0+t_{\rm var}$. The synchrotron emission from electrons has been corrected for synchrotron self absorption, which is important for the contribution of the cooling zone. In each panel, we show the value of the obtained fraction energy in protons divided by the total energy in electrons, $\Delta E_p/\Delta E_e$.  The very high-energy contributions shown ($p\gamma$, $pp$, and $e^+e^-$ synchrotron) are not corrected for $\gamma\gamma$ absorption in order to appreciate their intrinsic intensities. Also, since the redshift chosen for the example GRB is $z=1.8$, $\gamma\gamma$ annihilations of gamma-rays on the extragalactic background light (EBL) would cause complete absorption for $E_\gamma \gtrsim 100$ GeV  \cite[e.g.][]{inoue2012}.

Although we are not interested here in making predictions for the VHE photons and their detectability, for completeness we computed the synchrotron emission of a first generation of secondary $e^+e^-$ created by the decay of muons to verify that it does not overcome the synchrotron emission of the primary electrons, which is taken as the primary target for $p\gamma$ interactions. We obtained the corresponding distribution $N_{\mu\rightarrow e^\pm}$ as a solution of the kinetic equation with an injection taken to be equal to that of $\nu_e$, using the expression listed below after Lipari et al. (2007). Formally, a cascade will develop after internal $\gamma\gamma$ absorption, creating more pairs that will again radiate synchrotron photons and that can also get absorbed \citep[e.g.][]{asano2010}. While a complete treatment of such a cascade would give the final shape and intensity of the spectrum, we have checked that the synchrotron emission of the first generation of $e^+e^-$ pairs resulting from internal $\gamma\gamma$ absorption \citep{aharonian1983} is not greater than the emission from electrons and positrons from muon decays, as can be seen from Fig. \ref{fig6:E2dfluga_g100_g300}. Hence, assuming that after the full cascade the high energy part is reprocessed to lower energies, the expected intensity is not so high, and then the low energy photon field remains dominated by the synchrotron of primary electrons in each of the zones. This ensures that the neutrino output that we shall obtain from $p\gamma$ interactions is a good approximation in the cases studied.

We can point out a difference in the SED between the cases with different Lorentz factor. In the case of $\Gamma=100$, muons undergo acceleration because the magnetic field is higher, so the acceleration rate is greater than the decay rate. As a consequence, there is high synchrotron emission of these muons, unlike the case with $\Gamma=300$, for which the magnetic field is lower and there is no significant muon acceleration.
Other important difference between these two cases is clear through the values of the fraction $\Delta E_p/\Delta E_e$. As mentioned above, much more energy has to be present in protons in the case of $\Gamma=300$ in order to reach the same level of $p\gamma$ and $pp$ emissions as for $\Gamma=100$. This is because in the latter case, the corresponding cooling rates are much lower than the adiabatic cooling rate, making the proton emission processes less efficient.

\section{Neutrino emission}\label{sec:neutrinos}
Once we have the distributions of pions and muons, we can obtain the corresponding neutrino emissivities arising from their decay. The contribution to muon neutrinos and antineutrinos from the direct decay of pions is given by \cite[e.g.][]{lipari2007},
\begin{multline}
%\be
Q_{\pi\rightarrow\nu_\mu}(E,t)= \int_{E}^{\infty}dE_\pi
T^{-1}_{\pi,\rm d}(E_\pi)N_\pi(E_\pi,t) \\ \times
\frac{H(1-r_\pi-x)}{E_\pi(1-r_\pi)},
% \ee
\end{multline}
with $x=E/E_\pi$ and the decay timescale is $T_{\pi,\rm d}=2.6\times 10^{-8}{\rm s}$. 
The contribution from muon decays ($\mu^- \rightarrow e^- \bar{\nu}_e \nu_\mu$,  $\mu^+ \rightarrow e^+ {\nu}_e \bar{\nu}_\mu$) to muon neutrinos and antineutrinos is
and
\begin{multline}
%\be 
Q_{\mu\rightarrow\nu_\mu}(E,t)= \sum_{i=1}^4\int_{E}^{\infty}\frac{dE_\mu}{E_\mu} T^{-1}_{\mu,\rm
d}(E_\mu)N_{\mu_i}(E_\mu,t) \\ \times \left[\frac{5}{3}-
3x^2+\frac{4}{3}x^3 +\left(3x^2-\frac{1}{3}-\frac{8x^3}{3}\right)h_{i}
\right],
%\ee
\end{multline}
where $x=E/E_\mu$, $\mu_{1,2}=\mu^{-,+}_L$, $T_{\mu,\rm d}=2.2\times 10^{-6}{\rm s}$, and
$\mu_{3,4}=\mu^{-,+}_R$, and the helicity of the muons is $h=1$ for right-handed and $h=-1$ for left- handed muons. %\citep[see][]{2007PhRvD..75l3005L}.
Similarly, the emissivity of electron neutrinos and antineutrinos from the decay of muons is given by
\begin{multline}
Q_{\mu\rightarrow\nu_e}(E,t)= \sum_{i=1}^4\int_{E}^{\infty}\frac{dE_\mu}{E_\mu} T^{-1}_{\mu,\rm
d}(E_\mu)N_{\mu_i}(E_\mu,t) \\ \times \left[2-
6x^2+4x^3 +\left(2- 12x+ 18x^2-8x^3\right)h_{i}\right].
\end{multline}

The fluence obtained for a typical GRB is the sum of the contribution from both zones:
\be
\left. \frac{dN_{\nu_i}}{dE'_\nu}(E'_\nu)\right|_{\rm acc}&=&\int_{t>t_0}dt \ \mathcal{N}_{\rm inj}Q^{\rm acc}_{\nu_i}\left(E_\nu^{\rm (com)},t\right) \Delta V'\frac{dE_\nu^{\rm (com)}}{dE'_\nu} \nonumber \\
\left. \frac{dN_{\nu_i}}{dE'_\nu}(E'_\nu)\right|_{\rm cool}&=&\int_{t>t_0}dt \ \mathcal{N}_{\rm inj} Q^{\rm cool}_{\nu_i}\left(E_\nu^{\rm (com)},t\right) \Delta V' \frac{dE_\nu^{\rm (com)}}{dE'_\nu},\nonumber
\ee
where $Q^{\{\rm acc, cool\}}_{\nu_i}$ is the total $\nu_\mu +\bar{\nu}_\mu$ or $\nu_e +\bar{\nu}_e$ neutrino emissivity for the acceleration and cooling zones, in units (${\rm energy}^{-1}{\rm time}^{-1}{\rm length}^{-3}$); $E_\nu$ is the neutrino energy for $z=0$, the local neutrino energy is $E'_\nu=E_\nu(1+z)$, and the comoving one in the ejected flow is $E_\nu^{\rm (com)}\simeq E_\nu'/(2 \Gamma)$.

Considering the GRB redshift evolution rate \cite[e.g.][]{murase2006}
\beq 
 R_{\rm GRB}(z)= 23\frac{24\exp{(-3.05 \ z-0.4)}}{\exp (2.93 \ z)+ 15} \frac{\sqrt{\Omega_\Lambda+ \Omega_{\rm m}  (1+z)^3}}{(1+z)^{1/3}}\eeq
 in units of $ ({\rm Gpc}^{-3}{\rm yr}^{-1})$, the diffuse muon neutrino flux from GRBs can then be integrated in redshift:
\begin{multline}
  \Phi_{\nu_\mu}(E_\nu)= \frac{c}{4\pi H_0}\int_0^{z_{\rm max}}\frac{dz \ R_{\rm GRB(z)}}{\sqrt{\Omega_\Lambda+ \Omega_{\rm m}(1+z)^{3}}} \\ \times \left(\frac{dN_{\nu_\mu}\left[E_\nu(1+z)\right]}{dE'_\nu} {P_{\nu_{\mu}\rightarrow \nu_{\mu}}}+ \frac{dN_{\nu_e}\left[E_\nu(1+z)\right]}{dE'_\nu} {P_{\nu_{e}\rightarrow \nu_{\mu}}}\right), \label{dflunudiff}
\end{multline}
where $\Omega_\Lambda=0.7$, $\Omega_{\rm m}=0.3$, and $H_0=70 {\rm km \ s}^{-1} {\rm Mpc}^{-1}$. The effect of neutrino flavour oscillation is taken into account in Eq. (\ref{dflunudiff}) through the probability that the generated $\nu_\mu$ and $\bar{\nu}_\mu$ remain of the same flavour, $P_{\nu_{\mu}\rightarrow \nu_{\mu}}$, and also through the probability that electron neutrinos or antineutrinos and oscillate into muon neutrinos or antineutrinos, $P_{\nu_{e}\rightarrow \nu_{\mu}}$. These probabilities depend on the unitary mixing matrix $U_{\alpha j}$, which is determined by the three mixing angles $\theta_{12}\simeq 34^\circ$, $\theta_{13}\simeq 9^\circ$, and $\theta_{23}\simeq 45^\circ$, and a CP violating phase which we take to be zero. The values of these angles are derived from global fits to experimental data of solar, atmospheric, and accelerator neutrinos \cite[e.g.][]{gonzalezgarcia2012}, which yield the values for the probabilities $P_{\nu_{\mu}\rightarrow \nu_{\mu}}=0.369$ and
$P_{\nu_{e}\rightarrow \nu_{\mu}}\simeq 0.255$.   

In Fig. \ref{fig7:Esdflunu_g100_g300}, we show the different outputs for the background of muon neutrinos using $\Gamma=100$ and $\Gamma=300$, and with an escape-to-acceleration rate ratio of $\xi_{\rm esc}=0.25$ and $\xi_{\rm esc}=0.1$. For illustration, the parameters regulating the injected power ($L_e$ and $L_p$) and the efficiency of acceleration ($\eta$) have been chosen in order to obtain both a correct electron synchrotron emission (as compared to the typical broken power-law SED) and, at the same time, a neutrino flux at the level of the recent detection by IceCube \cite{klein2013,liu2013}. The effect of increasing $\eta$ would yield neutrinos that are more energetic than $\sim 10^6$ GeV and would also bring the electron synchrotron peak to higher energies, which would still be consistent with photon observations. If in light of new neutrino data \cite{aartsen2013} or clues disfavouring the association of the neutrino events with GRBs, it will be possible to exclude too high values of the injected power $L_p$ in the context of the present model.

\section{Discussion}

We have implemented a simple two-zone model in order to study the generation of high energy neutrinos associated with the prompt GRB emission. Using standard values for the magnetic field and size of the emission region, our model can account for the possible effect of the acceleration of secondary particles. In particular, we found that muons can efficiently gain energy if the magnetic field is strong enough, but still within attainable values in the context of GRBs. We note that these effects cannot be described with previous one-zone models that deal with neutrino emission in a magnetized environment \cite[e.g.][]{magnetic2009,baerwald2012}, in which the acceleration rate is only used to fix the maximum energy of the primary electrons and protons. 

As recognized in previous works \cite[e.g.][]{kirk1998}, particle acceleration can be accounted for using two zones and assuming that particles can escape from  the acceleration zone to the cooling zone. {We have not considered that particles in the cooling zone can further escape to a third zone in order not to miss their photon and neutrino output.} {The present model also differs from previous two-zone models in that the size of both zones are equal, and with a value derived from variability considerations. Including adiabatic losses for protons provides a mechanism for their faster cooling, on a timescale similar to the dynamical time, e.g. the one associated with the duration of the shell collision event in the internal shock scenario.  A variation of the present model could be implemented by including a convective term in the kinetic equation for the cooling zone \cite[e.g.][]{leptoh2011}. This would prevent us from having to impose a fixed size for the cooling zone, since particles of different species and energies would reach different distances as they cool.}

In the context studied here, we have found that if the escape rate is less than the acceleration rate ($\xi_{\rm esc}<1$), then the synchrotron emission from electrons in the acceleration zone dominates and can be the responsible for the usual GRB emission. Otherwise, for faster escape rates, the synchrotron emission from electrons in the cooling zone would dominate but with a spectrum too wide, which would greatly exceed the typical GRB emission at lower energies. As can be seen in Fig. \ref{fig6:E2dfluga_g100_g300}, for lower values of $\xi_{\rm esc}$, we obtain less significant electron synchrotron components from the cooling zone, and the bump corresponding to the acceleration zone falls within the correct energy range, as compared with the broken power-law benchmark.
By varying the acceleration efficiency of the different injection events (such as shell collisions in the internal shock model), different maximum energies for the electrons could be achieved, and their synchrotron emission would cover a window in the gamma-ray spectrum to be consistent with a full burst. In such cases with a low escape rate, we found that a neutrino component arising from the acceleration zone mainly by $p\gamma$ interactions becomes dominant at the highest neutrino energies, which in the examples shown reached $\sim 10^6$ GeV and can account for the recent IceCube data. 
 
 Some tasks could help make a more accurate calculation of the diffuse neutrino background in the context of the present type of models for GRBs: try to reproduce the observed gamma-ray spectrum of particular bursts by adjusting the number of acceleration events (peaks in the lightcurve) and the acceleration efficiency, and also to consider the probability of occurrence of bursts with different Lorentz factors. We leave these points for future work, along with the possible application of the model to other type of astrophysical sources.

\begin{acknowledgements}	
 I am especially thankful to Prof. A. Mastichiadis for many discussions, suggestions, and help, and I thank Stavros Dimitriakoudis and Maria Petropoulou for fruitful discussions on GRB physics. I also thank the referee P. M\'{e}sz\'{a}ros for a helpful review. Finally, I thank CONICET (Argentina) for their financial support and the University of Athens for their hospitality.
\end{acknowledgements}


\begin{thebibliography}{}
%   \bibitem[1980]{cox} Cox, J. P. 1980,
%      Theory of Stellar Pulsation
%      (Princeton University Press, Princeton) 165
%   \bibitem[1980]{mizuno} Mizuno H. 1980,
%      Prog. Theor. Phys., 64, 544
%   \bibitem[1992]{terlevich} Terlevich, R. 1992, in ASP Conf. Ser. 31, 
%      Relationships between Active Kardashev, N. S.
%      ed. A. V. Filippenko, 13
%\bibitem{blandford1976} R.~D. Blandford and C.~F. McKee, {\it Phys. Fluids}, {\bf 19}, 1130 (1976)

\bibitem[Aartsen et al. 2013]{aartsen2013} Aartsen, M. G., et al., IceCube Collaboration 2013, Science 342, 6161
%\bibitem[Abassi et al. 2010]{abbasi2010}  Abbasi, R., et al., IceCube Collaboration 2010, ApJ 710, 346
\bibitem[Aharonian et al. 1983]{aharonian1983} Aharonian, F. A., Atoyan, A. M., \& Nagapetian, A. M. 1983, Astrofizika {19}, 323
\bibitem[Ando \& Beacom 2005]{ando2005}Ando, S. \&  Beacom, J. F. 2005, PRL 95, 061103
\bibitem[Asano et al. 2010]{asano2010}
  Asano, K., Inoue, S., \& Meszaros, P. 2010, ApJL, 725, L121
  %``Prompt X-ray and Optical Excess Emission due to Hadronic Cascades in Gamma-Ray Bursts,''
%  arXiv:1009.5178 [astro-ph.HE].
\bibitem[Atoyan \& Dermer 2003]{atoyandermer2003} Atoyan, A. M. \& Dermer, C. D. 2003, ApJ {586}, 79
\bibitem[Baerwald et al. 2012]{baerwald2012}
 Baerwald, P., H\"{u}mmer, S. \& Winter,  W. 2012,
Astropart. Phys. {35}, 508
\bibitem[Begelman et al. 1990]{begelman1990}Begelman, M. C., Rudak, B., \& Sikora, M. 1990, ApJ, 362, 38
%\bibitem{km3net} Bersani, A. for the KM3NeT Consortium, 2012 AIP Conf. Proc. 1441, 429
%Electromagnetic extraction of energy from Kerr black holes
\bibitem[Blumenthal \& Gould 1970]{blumenthal1970} Blumenthal, G. R., \& Gould, R. J. 1970, Rev. Mod. Phys., 42, 237
\bibitem[Bosch-Ramon 2012]{boschramon2012} Bosch-Ramon, V. 2012, A\&A 542, A125 
\bibitem[Dai \& Lu 2001]{dai2001}Dai, Z. G., Lu, T. 2001, ApJ 551, 249
\bibitem[Drury et al. 1999]{drury1999} Drury, L., Duffy, P.,  Eichler, D.,  \& Mastichiadis,  A. 1999, A\&A {347}, 370 
\bibitem[Drury 2012]{drury2012} Drury, L. 2012, MNRAS 422, 2474
\bibitem[Fenimore et al. 1996]{fenimore1996}Fenimore, F. E., Madras, C. D., \& Nayakshin, S. 1996, ApJ 473, 998
\bibitem[Fletcher et al. 1994]{sibyll1994} Fletcher, R. S., Gaisser, T. K., Lipari, P., \& Stanev, T. 1994, Phys.
Rev. {D 50} 5710
\bibitem[Gao et al. 2011]{gao2012}
  Gao, S., Asano, K. \& Meszaros, P.
  %``High Energy Neutrinos from Dissipative Photospheric Models of Gamma Ray Bursts,''
  2011, JCAP {1211}, 058
%  [arXiv:1210.1186 [astro-ph.HE]].
%\bibitem{dominguez2010} Dom\'{\i}nguez, A. et al. 2010 {MNRAS}, {410}, 2556
%\bibitem{ferrari1981}A. Ferrari , E. Trussone, and L. Zaninetti, {\it
%MNRAS}, {\bf 196}, 1051 (1981)
%\bibitem{Giroletti:2003wt}
%  Giroletti, M., {et al.} 2004, ApJ {600} 127
  %``Parsec scale properties of Markarian 501,''
%\bibitem{ghisellini2002}Ghisellini, G., Celotti, A., \& Costamante, L. 2002, A\&A, 386, 833
%\bibitem{ghisellini2005}Ghisellini, G., Tavecchio, F., \& Chiaberge, M. 2005 {A\& A}, {432}, 401
%\bibitem{Kappes:2006fg}
%Kappes, A., Hinton, J., Stegmann, C., \& Aharonian, F.~A.
%  %``Potential Neutrino Signals from Galactic Gamma-Ray Sources,''
%  2007, ApJ, {656}, 870
%   [Erratum-ibid.\ 2007, {661}, 1348]
%\bibitem{Kobayashi:1997jk} Kobayashi, S., Piran, T.,  \& Sari, R. 1997, ApJ, 490, 92
  %``Can internal shocks produce the variability in GRBs?,''
%	Launching of jets by cold, magnetized disks in Kerr metric
%\bibitem{}Schlickeiser, R. \& Lerche, I. 2007, A\&A, 476, 1


\bibitem[Giannios 2006]{giannios2006} 
Giannios, D. 2006,
  %``Prompt emission spectra from the photosphere of a grb,''
  A\&A {457}, 763
%  [AIP Conf.\ Proc.\  {\bf 924}, 32 (2007)]
 % [astro-ph/0602397].
\bibitem[Gonzalez-Garc\'{\i}a et al. 2012]{gonzalezgarcia2012} 
 Gonzalez-Garc\'{\i}a, M. C, Maltoni, M., Salvado, J., \& Schwetz, T.
  %``Global fit to three neutrino mixing: critical look at present precision,''
 2012,  JHEP {1212}, 123
\bibitem[Guetta et al. 2004]{guetta2004} Guetta, D., Hooper, D., Alvarez-Muñiz, J., Halzen, F.,
\& Reuveni, E. 2004, Astropart. Phys. 20, 429 
\bibitem[H\"{u}mmer et al. 2012]{hummer2012} H\"{u}mmer, S., Baerwald, P. \& Winter, W. Phys. 2012, Rev. Lett. 108, 231101


\bibitem[He et al. 2012]{he2012} 
  He, H.~-N., Liu, R.~-Y., Wang, X.~-Y., Nagataki,  S., Murase,  K., and Dai, Z.~-G.
  %``Icecube non-detection of GRBs: Constraints on the fireball properties,''
  ApJ.\  {752}, 29 (2012)


\bibitem[Inoue et al. 2012]{inoue2012}
  Inoue, Y., Inoue, S., Kobayashi, M.~A.~R., Makiya, R., Niino, Y. \& ~Totani, T. 2013,
  %``Extragalactic Background Light from Hierarchical Galaxy Formation: Gamma-ray Attenuation up to the Epoch of Cosmic Reionization and the First Stars,''
  ApJ {768}, 197
%  [arXiv:1212.1683 [astro-ph.CO]].

\bibitem[Jones 1968]{jones1968} Jones, F. C. 1968, Phys. Rev. 167, 1159
\bibitem[Kardashev 1962]{kardashev1962} Kardashev, N. S. 1962, Soviet Astronomy 6, 317
%\bibitem{kazanas2002}
%  Kazanas, D., Georganopoulos, M., \& Mastichiadis, A.
  %``The Supercritical pile model for GRB: Getting the nu F(nu) peak at 1 MeV,''
 %2002, Astrophys.\ J.\  {578}, L15
\bibitem[Kelner et al. 2006]{kelner2006}Kelner,  S. R., Aharonian, F. A., \&  Bugayov, V. V. 2006, Phys. Rev. {D 74}, 034018
\bibitem[Kirk et al. 1998]{kirk1998}Kirk, J. G., Rieger, F. M., Mastichiadis, A. 1998, A\&A 333, 452
\bibitem[Klein 2013]{klein2013}Klein, S., Highlight talk at ICRC 2013, Jul 4th 2013, Rio de
Janeiro, Brazil

\bibitem[Kobayashi et al. 1997]{kobayashi1997}Kobayashi, S., Piran, T., \& Sari, R. 1997, ApJ 490, 92
\bibitem[Lipari et al. 2007]{lipari2007} Lipari, P., Lusignoli, M., \& Meloni, D. 2007, Phys. Rev. {D 75}, 123005
\bibitem[Liu et al. 2013]{liu2013}
Liu, R.-Y., Wang, X.-Y., Inoue, S., Crocker, R., \& Aharonian, F.
	  2013, [arXiv:1310.1263 [astro-ph.HE]].
\bibitem[M\'{e}sz\'{a}ros 2006]{meszaros2006}
  M\'{e}sz\'{a}ros, P. 2006,
  %``Gamma-Ray Bursts,''
  Rept.\ Prog.\ Phys.\  {69}, 2259
%  [astro-ph/0605208].
\bibitem[M\'{e}sz\'{a}ros \& Rees 2006]{meszaros2000}  M\'{e}sz\'{a}ros, P. \&  Rees, M. J. 2000, ApJ 530, 292
% arXiv:astro-ph/9908126.
\bibitem[Moraitis \& Mastichiadis 2007]{moraitis2007}Moraitis, K. \& Mastichiadis,  A. 2007, A\&A 462, 173 
\bibitem[Murase \& Nagataki 2006]{murase2006} Murase, K., \& Nagataki, S. 2006, Phys. Rev. {D 73}, 063002 
\bibitem[Murase 2007]{murase2007} Murase, K. 2007, Phys. Rev. {D 76}, 123001
\bibitem[Murase et al. 2012]{murase2012} Murase, K., Asano,  K., Terasawa, T., \& M\'{e}sz\'{a}ros, P.
 2012,  ApJ {746}, 164 
\bibitem[M\"{u}cke et al. 2000]{sophia2000}  
  M\"{u}cke, A., et at. 2000, Comp. Phys. Comm. {124}, 290 
\bibitem[Piran 2004]{piran2004}
  Piran, T.
  %``The physics of gamma-ray bursts,''
 2004, Rev.\ Mod.\ Phys.\  {76}, 1143
\bibitem[Protheroe \& Stanev 1999]{protheroestanev1999} Protheroe, R. J. \&  Stanev, T. 1999, Astropart. Phys. {1O}, 185

\bibitem[Razzaque et al. 2004]{razzaque2004} Razzaque, S.,  Meszaros, P., \&  Waxman, E. 2004, Phys. Rev.
Lett. 93, 181101; 94, 109903(E) (2005).

\bibitem[Rees \& M\'{e}sz\'{a}ros 1994]{rees1994}Rees, M. J. \& M\'{e}sz\'{a}ros, P. 1994, ApJ, 430, L93

\bibitem[Reynoso \& Romero 2009]{magnetic2009}Reynoso, M. M. \& Romero, G. E. 2009, A\&A, 493, 1
\bibitem[Reynoso et al. 2011]{leptoh2011} 
Reynoso, M. M., Medina, M. C., \& Romero, G. E. 2011, A\&A,  531,  A30 
  %``A lepto-hadronic model for high-energy emission from FR I radiogalaxies,''
\bibitem[Vieyro et al. 2013]{vieyro2013}Vieyro, F. L., Romero, G. E., Peres, O. L. G. 2013, A\&A 558, A142
\bibitem[Waxman \& Bahcall 1997]{waxmanbahcall1997}
 Waxman, E. \& Bahcall, J. 1997, Phys. Rev. Lett. {78}, 2292
 \bibitem[Waxman \& Bahcall 2000]{waxman2000}
  Waxman, E. \& Bahcall, J. N. 2000,
  %``Neutrino afterglow from gamma-ray bursts: Similar to 10**18-eV,''
  ApJ {541} 707
  %  [hep-ph/9909286].
  %%CITATION = HEP-PH/9909286;%%
\end{thebibliography}
\end{document}